\documentclass[compsoc,conference,a4paper,10pt,times]{IEEEtran}
\IEEEoverridecommandlockouts

\usepackage{pgf-umlsd}
\usepackage{tikz}
\usepackage{amsmath}
\usepackage{tkz-graph}
\usepackage{xspace}
\usepackage{amsthm}
\usepackage{anyfontsize}
\usepackage{pifont}
\usepackage{url}
\usepackage{bold-extra}
\usepackage{fancyhdr}
\usepackage[ruled, vlined,linesnumbered,noend]{algorithm2e}
\usepackage{enumitem}

\usepackage{cite}
\usepackage{algorithmic}
\usepackage{graphicx}
\usepackage{lipsum}
\newcommand{\subparagraph}{}

\usepackage{titlesec}

\titlespacing\section{0pt}{4pt plus 2pt minus 2pt}{2pt plus 2pt minus 2pt}
\titlespacing\subsection{0pt}{4pt plus 2pt minus 2pt}{2pt plus 2pt minus 2pt}
\titlespacing\subsubsection{0pt}{4pt plus 2pt minus 2pt}{2pt plus 2pt minus 2pt}

\usepackage[colorlinks=true,urlcolor=black]{hyperref}
\def\BibTeX{{\rm B\kern-.05em{\sc i\kern-.025em b}\kern-.08em
    T\kern-.1667em\lower.7ex\hbox{E}\kern-.125emX}}

\usepackage{breakurl}

\usepackage{bm}

\makeatletter
\renewcommand{\paragraph}{%
  \@startsection{paragraph}{4}%
  {\z@}{1.2ex \@plus 1ex \@minus .2ex}{-1em}%
  {\normalfont\normalsize\bfseries}%
}
\makeatother

\newcommand{\bnegspace}{\vspace{-1.5\baselineskip}}

\newcommand{\negspace}{\vspace{-0.5\baselineskip}}
\newcommand{\snegspace}{\vspace{-0.25\baselineskip}}

\SetAlFnt{\scriptsize\sffamily}

\usepackage[colorinlistoftodos]{todonotes}
\newcounter{todocounter}

\newtheorem{theorem}{Theorem}

\newtheorem{lemma}{Lemma}
\usepackage{amsthm}
\let\oldproofname=\proofname
\renewcommand{\proofname}{\rm\bf{\oldproofname}}

\newcommand{\sys}{Ostraka\xspace}

\newcommand{\shard}{\textit{NS}\xspace}
\newcommand{\shards}{{\shard}s\xspace}

\newcommand{\subc}{subchain\xspace}
\newcommand{\subcs}{{\subc}s\xspace}

\newcommand{\bshard}{{block-shard}\xspace}

\newcommand{\salt}{{salt}\xspace}
\newcommand{\salts}{\textit{salts}\xspace}
\newcommand{\txhash}{\texttt{TxHash}\xspace}
\newcommand{\newtxhash}{\texttt{NewTxHash}\xspace}
\newcommand{\msb}{\textit{Df}\xspace}
\newcommand{\pow}{\textit{PoW}\xspace}
\newcommand{\pos}{\textit{PoS}\xspace}
\newcommand{\dos}{{DoS}\xspace}
\newcommand{\bad}{\textsc{BadBlock}\xspace}
\newcommand{\inv}{\textsc{Inventory}\xspace}

\begin{document}

\newcommand{\mem}{mempool\xspace}
\newcommand{\utxoset}{{UTXO set}\xspace}
\newcommand{\utxoshard}{{UTXO-shard}\xspace}

\newcommand{\sinter}[2]{\ensuremath{S_{#1}^{#2}\xspace}}
\newcommand{\sintra}[1]{\ensuremath{S_{#1}}\xspace}
\newcommand{\nod}[1]{\ensuremath{N^{#1}}\xspace}
\newcommand{\cod}[1]{\ensuremath{C^{#1}}\xspace}
\newcommand{\blk}{\ensuremath{B}\xspace}
\newcommand{\sid}{\textit{ShardID}\xspace}
\newcommand{\sha}{\ensuremath{\textsc{SHA-256}}\xspace}

\date{}

\title{\Large \bf \sys: Secure Blockchain Scaling by Node Sharding}

\author{\IEEEauthorblockN{Alex Manuskin}
\IEEEauthorblockA{\textit{Technion, IC3} \\
amanusk@pm.me}
\and
\IEEEauthorblockN{Michael Mirkin}
\IEEEauthorblockA{\textit{Technion, IC3} \\
michael.mirkin@gmail.com}
\and
\IEEEauthorblockN{Ittay Eyal}
\IEEEauthorblockA{\textit{Technion, IC3} \\
ittay@technion.ac.il }
}

\maketitle

\sloppy

\begin{abstract}
Cryptocurrencies, implemented with blockchain protocols, promise to become a global payment system if they can overcome performance limitations. 
Rapidly advancing architectures improve on latency and throughput, but most require all participating servers to process all transactions. 
Several recent works propose to \emph{shard} the system, such that each machine would only process a subset of the transactions. 

However, we identify a denial-of-service attack that is exposed by these solutions -- an attacker can generate transactions that would overload a single shard, thus delaying processing in the entire system. 
Moreover, we show that in common scenarios, these protocols require most node operators to process almost all blockchain transactions. 

We present Ostraka, a blockchain node architecture that shards (parallelizes) the nodes themselves. 
We prove that replacing a unified node with an Ostraka node does not affect the security of the underlying consensus mechanism.

We evaluate analytically and experimentally block propagation and processing in various settings.
Ostraka allows nodes in the network to scale, without costly coordination.
In our experiments, Ostraka nodes transaction processing rate grows linearly with the addition of resources.
\end{abstract}

\begin{IEEEkeywords}
Distributed systems, Blockchain, Network security
\end{IEEEkeywords}

    \section{Introduction}
    \label{sec:introduction}

    Payment systems are typically controlled by trusted entities~-- central banks control supply by printing money, banks transact among each other with a central national~\cite{spfs} or international~\cite{scott2012origins} mechanism, and credit card companies~\cite{visa56ktxs} authorize their clients' transactions.
This centralization of power implies limitations of the users' freedom~\cite{riksbank2017eKrona}, less resilience, and high fees~\cite{ausubel1991failure}.

Since their introduction with Bitcoin~\cite{nakamoto2008bitcoin}, cryptocurrencies brought the promise of a \textit{decentralized} global payment system outside the control of a single entity.
Such a system is potentially useful for payment among individuals, but also among banks, machines (IoT), and micro-payments for social network interactions.


So far, blockchains systems~-- the decentralized mechanism on which cryptocurrencies operate~-- could not reach rates desired for global payment systems.
Early system implementations suffered from a consensus bottleneck, limiting them to less than a dozen transactions per second~\cite{wood2014ethereum} and they cannot scale by tuning their parameters~\cite{croman2016scaling}. 

More recent protocols~\cite{cachin2016architecture,yin2019hotstuff,ng2016eyal,solidus2016abraham,algorand2017gilad,byzcoin2016kogias,ouroboros2017kiayias,rocket2018snowflake,bagaria2019prism} overcome the consensus bottleneck, but they are limited by the processing capacity of the nodes comprising the system.
We provide background in~\S\ref{sec:preliminaries}. 

Each node is required to store and process each block broadcasted on the network.
As the number of transactions grows, and the longer the blockchain exists, it becomes increasingly difficult for nodes to keep up.
Existing designs, where each node is operated on a single machine are not equipped to scale with the rising demand.

Several recent solutions~\cite{luu2016secure,omniledger2017kogias,rapidchain2018zamani,al2017chainspace} propose to \emph{shard} the blockchain, splitting it into multiple interleaved \subcs. 
Each \subc is maintained by a distinct set of nodes, and so each node does not have to process all transactions in the system but only those in its \subc. 
This allows the system to process transactions at a higher rate while reducing the load on the individual node. 

Those solutions target \emph{democratic environments}, where there are many participants with similar resources. 
In such environments, increasing the number of participants allows to increase the number of \subcs and thus system capacity. 
However, we show that when resource distribution is not democratic, as is the case in most popular cryptocurrencies, the security and performance of those systems deteriorate. 
We show how they could be vulnerable to attacks, and why they do not necessarily provide the benefits of lowering the burden on the single node.
We discuss these solutions and related work in~\S\ref{sec:related}. 

In this work, we present \emph{Ostraka}, a novel architecture that allows for linear scaling of blockchain \emph{nodes} capacity to meet network requirements.
Ostraka enables scaling without forfeiting security assumptions.
Rather than partitioning the system's state into separately-managed clusters, in \sys we scale the nodes themselves, splitting each node across multiple \emph{Node-Shards (\shards)}, allowing vertical scaling of each node.
The challenge is creating a partitioning scheme that achieves several goals.
Data must be accessible in~$O(1)$ while keeping the system secure and preventing an adversary from slowing down the validation process across the network.
We describe the architecture in~\S\ref{sec:architecture}. 

The implication is that nodes require more resources, so fewer entities will run such nodes (e.g hundreds instead of thousands currently in Bitcoin~\cite{coindance2019} and Ethereum~\cite{eth2019explorer}). 
We argue, however, that the same is true for \subc based proposals, where only an optimistic \emph{democratic} setup would lead to better decentralization.
In more realistic setups, with similar properties to those in active systems today, they would also imply a smaller number of full participants.
Importantly, the decrease in node count does not mean that the system becomes closed: Anyone can still join the system and participate in the protocol.
Meanwhile, as with current cryptocurrencies, end-users who send and receive payments can still use their mobile devices by querying a few nodes.

\sys is a node architecture and is thus independent of the underlying consensus protocol. 
We achieve improved throughput by applying a novel block validation algorithm, that allows even distribution of all aspects of block validations across multiple machines~(\S\ref{sec:architecture}).
We demonstrate how we achieve these properties, without formfitting existing security guarantees.

We prove that the security guarantees of the protocol are the same as those of a theoretical high-performance \emph{unified} (standard) node~(\S\ref{sec:security}).
In contrast to existing solutions~\cite{nakamoto2008bitcoin, ethereum2015white}, Ostraka does not regard transaction order within each block. 
This avoids expensive locking mechanisms requiring elaborate communication. 
We prove the correctness of this crucial enabler of distributed execution and show that the Ostraka distributed node is operationally indistinguishable from a unified-node, despite order changes and processing parallelization.
Next, we show that an attacker can only perform a denial-of-service (DoS) attack against a small number of \sys nodes with high probability.

We evaluate Ostraka's block propagation in a variety of configurations and network conditions.
We show that an \sys node's processing rate grows nearly optimally with the number of \shards.
We provide theoretical analysis for the node performance as a function of inter- and intra-node bandwidth.
We measure the performance of \sys in transaction processing with as many as~64 \shards, achieving a rate of nearly 400$k$ transactions per second. 
We demonstrate the effect of Ostraka on a full system through simulation of Bitcoin-NG~\cite{ng2016eyal}, achieving an order of magnitude improvement. 
Evaluation details are in~\S\ref{sec:evaluation}. 

In summary, our main contributions are:
\begin{itemize}
\item Analysis of previous scaling solutions in non-democratic environments; 
\item Ostraka -- a novel blockchain scaling architecture; 
\item Proof that Ostraka does not reduce security compared to a unified node architecture; 
\item Theoretical performance analysis showing linear scaling -- up to network limits; 
\item Node implementation performance matching theory; 
\item A full system simulation. 
\end{itemize}

    \section{Preliminaries: The UTXO Model}
    \label{sec:preliminaries}

Ostraka utilizes the UTXO model~\cite{antonopoulos2014mastering,narayanan2016bitcoin,bitcoinwiki} that was introduced in Bitcoin and is used in many systems.
We describe here the elements relevant to this work.

\negspace
\paragraph*{Transaction}
Each transaction comprises two lists~-- inputs and outputs.
The outputs are the new records indicating ownership of amounts.
Each transaction is uniquely identified by a hash of its contents, ~\txhash.
An input of a transaction references an output it \emph{spends} by the~\txhash of the transaction containing the output and its index in its output list.
To spend an output, a spending condition must be satisfied, usually providing a cryptographic signature.

\paragraph*{Node data structures}
To operate a blockchain system, each of the nodes maintains three data structures: 
a set of transactions that were not yet placed in the blockchain~-- the \textit{mempool};  
the blockchain itself, a series of blocks containing transactions; and
the current system state~-- the UTXO set.

Users create transactions and broadcast them, and the nodes store them in their mempools. 
Once a \emph{miner} constructs a block from transactions in its mempool, she publishes the block to its peers.
When the node receives a new block, it removes all outputs spent by transactions in the block from the \utxoset and adds the newly created outputs.
The state of the \utxoset is thus the state of the blockchain following each block.

\negspace
\paragraph*{Security in \textit{Proof-of-Work} Blockchains}
In a \pow blockchain, block propagation time directly affects system security.
Nakamoto consensus states that the most up-to-date state of the blockchain is the chain with the most work (the longest chain).
If block propagation is slow, it takes time until other miners become aware of the new block. 
So two (or more) miners might generate blocks with the same predecessor. 
This split is called a \textit{fork}.

To prevent forks, block propagation time must be much shorter than the interval between blocks.
However, if nodes do not validate blocks before propagating them further, the system could be susceptible to \dos attacks, by flooding the system with invalid blocks.
Reducing block processing time is therefore critical to blockchain security~\cite{decker2013information,croman2016scaling,gencer2018}. 

\begin{figure*}[!htb]
\centering
\begin{minipage}{.45\textwidth}
  \centering
    \includegraphics[width=1.0\linewidth]{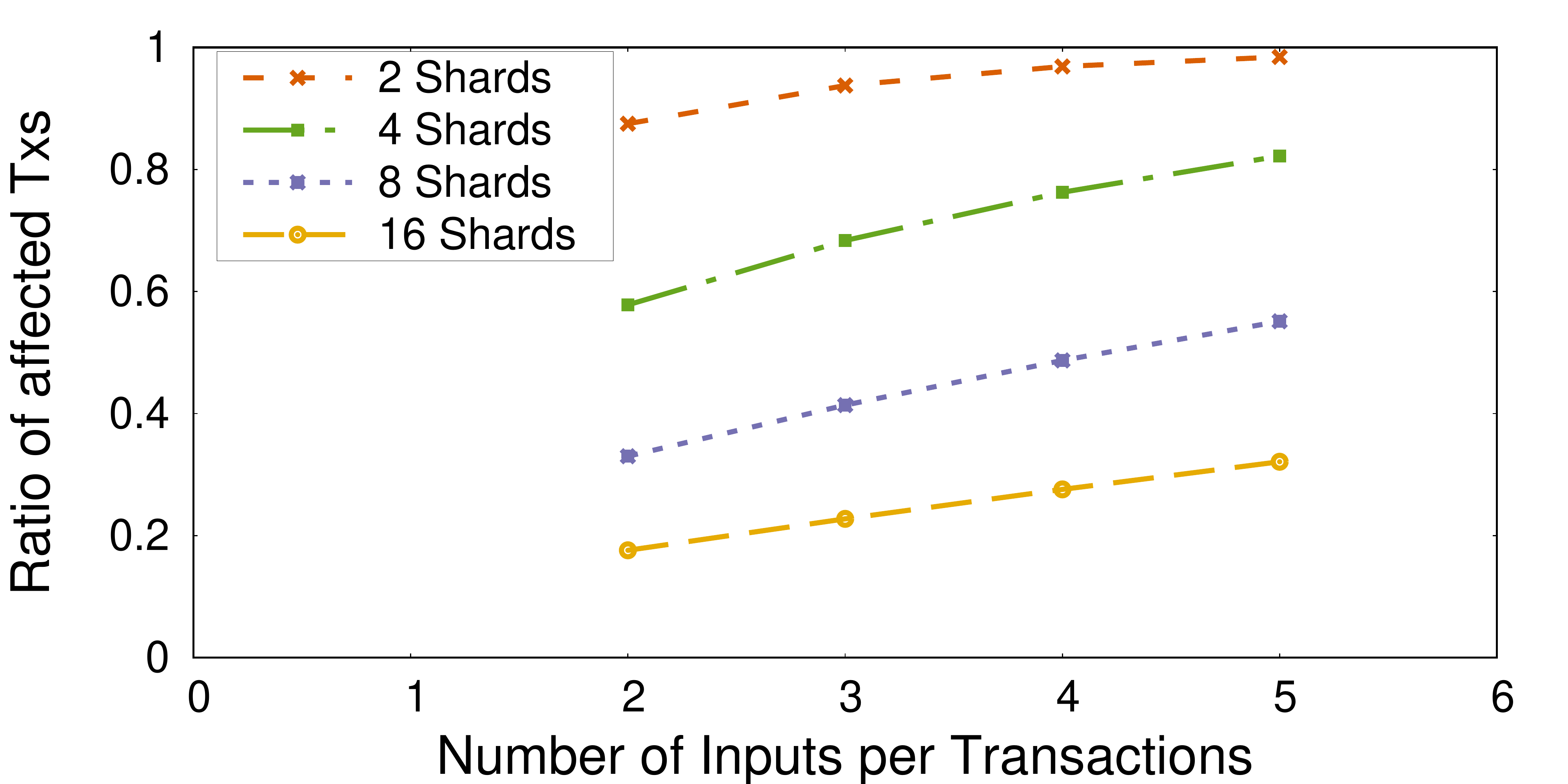}
    \caption{Ratio of effected transactions by attacking a single \subc}
    \label{fig:effected}
\end{minipage}
\hfill
\begin{minipage}{.45\textwidth}
  \centering
    \centering
    \includegraphics[width=1.0\linewidth]{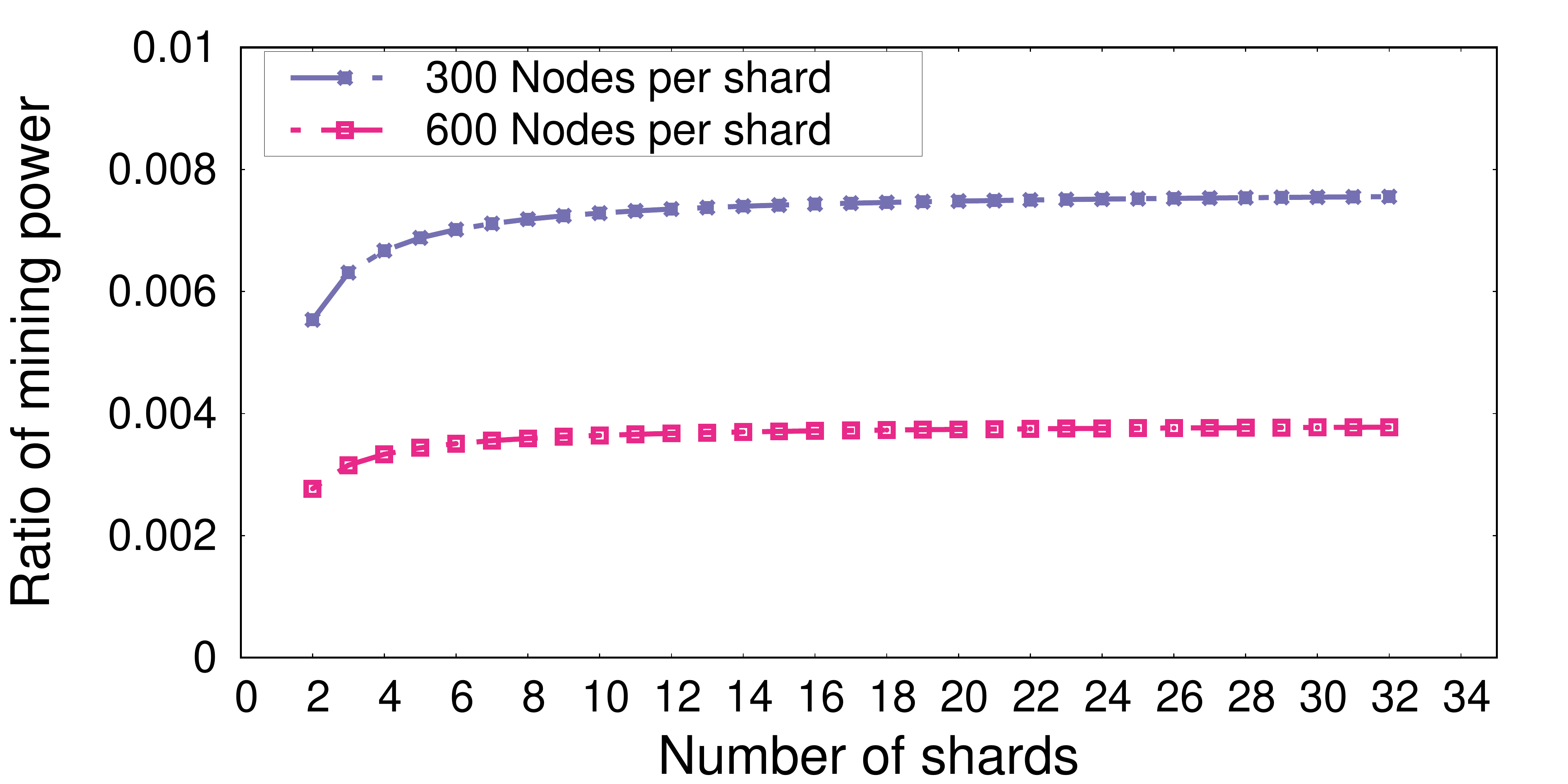}
    \caption{Ratio of mining power requiring participation in all \subcs}
    \label{fig:shares}
\end{minipage}%
\end{figure*}


    \section{Sub-chain Analysis and Related Work}
    \label{sec:related}

We overview current approaches to scaling blockchain performance, starting with novel security and performance analysis of proposals to split the system into \subcs. 


        \subsection{Concurrent \subcs (Sharding)}
With this approach, the system is divided into serveral \subcs.
Each \subc is maintained by a committee running a Byzantine Fault Tolerant (BFT) protocol, tolerant of an adversary of up to 1/3 of the network.
Elastico~\cite{luu2016secure} was one of the first proposals of this scheme.
OmniLedger~\cite{omniledger2017kogias} uses \subcs and builds on top of ByzCoin~\cite{byzcoin2016kogias}.
RapidChain~\cite{rapidchain2018zamani} closely follows the ideas presented in Elastico and OmniLedger but proposes different solutions to committee election and cross-\subc transactions, not dependent on the user.
Both OmniLedger and RapidChain use the UTXO model to distribute transactions across \subcs.
Chainspace~\cite{al2017chainspace} suggests a similar solution for a smart contracts system~\cite{wood2014ethereum}. 
\negspace
\paragraph*{DoS vulnerability} 
These systems can be vulnerable to DoS attacks.
Transactions are assigned to \subcs according to their \txhash.
An attacker can overwhelm a single \subc with many transactions much more cheaply than saturating the entire system. 
Other \subcs are also throttled due to transactions that originate from multiple \subcs. 

We evaluate the effectiveness of this attack. 
Consider a system of~$n$ \subcs and transactions with a single output and~$k$ inputs. 
The probability of a transaction to either have its output or one of its inputs in the \subc undergoing the attack is the complement of the probability that the output and all inputs not in the attacked \subc, $1 - \left( \frac{n - 1}{n} \right)^{k+1}$. 
Note that this is a lower bound: The number of transactions not able to execute is even higher when considering transaction dependant on pending transactions.
Fig.~\ref{fig:effected} illustrates this result. It shows, for example, that even with~2-3 inputs, in a 2-\subc system about~$90\%$ of transactions are affected, but even with~16 \subcs about~$20\%$ of the transactions are affected. 

\negspace
\paragraph*{Performance} 

Sub-chain systems use various mechanisms for cross-\subc transactions.
OmniLedger~\cite{omniledger2017kogias} uses 2-phase commit; 
it requires the client to obtain proofs of spending from each \subc where she wants input to be spent.
RapidChain~\cite{rapidchain2018zamani} relies on the leader of the \emph{output} \subc to create new outputs, mirroring outputs spent on \emph{input} \subcs.
In both approaches, since most transactions affect multiple \subcs, the additional overhead is considerable.
The architecture of Ostraka avoids this limitation.

\negspace
            \paragraph*{Node load} 
Ostraka nodes distribute the data internally, such that every node maintains the entire state. 
This is seemingly a difference from \subc systems where each node only maintains a subset of the state. 
However, as we show below, this difference only holds in what we call \emph{democratic environments}, where there are many miners (stakeholders for PoS) with similar mining power~\cite{omniledger2017kogias,luu2016secure,rapidchain2018zamani}. 
In non-democratic environments, where the number of miners is small, \subc systems also require nodes to store the entire state. 
Prominent systems relying either on \pow or \pos are typically non-democratic~\cite{arewedecent, gencer2018}, and recent findings~\cite{kwon2019impossibility} shows that indeed true decentralization is not possible in any Sybil protection mechanism used today.
As of today, in Bitcoin, for example, about~85\% of the mining power is controlled by~10 mining entities~\cite{mining-dist}.

Recall that \subc systems rely on BFT protocols to maintain each of the \subcs. 
The underlying BFT protocol assumes a bound on the number of Byzantine nodes, where a node is a unit of mining power.
Even if the bound holds for the entire system (e.g., less than a third of the mining power belongs to malicious principals), in order to guarantee the bound, there should be sufficiently many nodes in each \subc~\cite{omniledger2017kogias}. 
With a 25\% adversary, each \subc should have at least 600 nodes~\cite{omniledger2017kogias}.
Studies~\cite{gencer2018} show that \pow systems are as decentralized as 20 nodes operating in a quorum, a far cry from the hundreds required for good security.
Since the number of principals is not necessarily this large, participants instead split their mining power such that every mining power unit allows its owner one virtual node or \emph{share} in the system. 
To prevent an attacker from targeting a specific \subc, each share is assigned to a \subc uniformly at random. 

We estimate the minimal miner size for which a miner will have shares in most of the \subcs. 
Denote by~$S$ the number of nodes per \subc and by~$n$ the number of \subcs. 
Let~$r$ be the target ratio of \subcs that should be occupied in expectation and denote the miner size to achieve the target by~$\alpha$. 
The number of shares belonging to the miner is~$\alpha n S$. 
The expected number of \subcs occupied by the miner is therefore  
$n \left( 1 - \left( \frac{n - 1}{n} \right)^{\alpha n S} \right)$. 
This expression should equal the target number of \subcs, $r n$, and solving for~$\alpha$ we get 
$\alpha = \frac{\log(1 - r)}{S n \log(1 - 1/n)}$. 

To obtain concrete numbers, we take~$S$ to be either~300 or~600 as required by~Kogias et al.~\cite{omniledger2017kogias} and vary the number of \subcs,~$n$, from~1 to~32. We target an expected~$90\%$ of the \subcs and obtain the results shown in Fig.~\ref{fig:shares}. 
Indeed, even with only 300 nodes per \subc and 32 \subcs, a miner would participate in~$90\%$ of the \subcs in expectation if its size is larger than~$0.8\%$

The implication is that in non-democratic scenarios, which are common in active blockchains, the miners must participate in all \subcs, requiring them to process the entire chain, therefore not achieving the goal of reducing stress on the single node.

\subsection{Node scaling proposals}
Dickerson et al.~\cite{dickerson2017adding}, implement concurrent execution in smart contracts in a virtual machine model as used in Ethereum~\cite{wood2014ethereum}. 
They add a concurrent schedule to each block, enabling consistent parallel execution across different nodes. 
However, the nature of the VM model limits the level of their measured speed up to 2x.
Read-after-Write dependencies of instructions in the VM model enforce a sequential execution to avoid an inconsistent state.
Ostraka does not suffer from these limitations; we observed an improvement of two orders of magnitude, limited only by the network and the capacity of our experimental testbed. 

\negspace 
    \section{Architecture}
    \label{sec:architecture}

\newcommand{\view}{\textit{view}\xspace}
\newcommand{\tx}[1]{\ensuremath{\textit{TX}_{#1}\xspace}}
\newcommand{\inp}[1]{\ensuremath{\textit{IN}_{#1}\xspace}}
\newcommand{\inps}{\textit{Ref\_Out}\xspace}
\newcommand{\inpr}{\textit{Requested\_Out}\xspace}
\newcommand{\utxon}[1]{\ensuremath{\textsc{UTXO\_set}_{#1}\xspace}}
\newcommand{\empt}{\ensuremath{\bot}\xspace}
\newcommand{\txo}{\ensuremath{\textit{TX\_out}}\xspace}

\newcommand{\out}[1]{\ensuremath{\textit{OUT}_{#1}\xspace}}

We proceed to describe the structure and operation of \sys.
An \sys node operates similarly to a unified node. 
The key difference is that in \sys the \mem, \utxoset, and blockchain storage are distributed among several machines, called \textit{node-shards} (\shards). 

We describe the system components~(\S\ref{sub:arch_system_componenets}), explain the techniques used to make it efficient and secure~(\S\ref{sub:arch_sesign}), and overview system operation~(\S\ref{sub:arch_operation}).

        \subsection{System Components}
        \label{sub:arch_system_componenets}

In \sys a \textit{coordinator} keeps track of the blockchain and orchestrates communication within the node (intra-node) and with other nodes (inter-node).
We refer to the \shards of the same node as \textit{sibling \shards}.
The coordinator can operate as a separate machine or as a process on one of the \shards.
In \sys work distribution is local on each node and does not mandate the same number of \shards per node.
Transactions are distributed among \shards as atomic units.
Transaction outputs are stored in the \emph{UTXO-shard} of the \shard they are assigned to.
Note we assume complete trust among the \shards and the coordinator within a node. 

\negspace
\paragraph*{Coordinator}
The coordinator orchestrates the node operation.
When the node is first initialized, all \shards connect to the coordinator. 
The coordinator coordinates communication between sibling \shards and with other nodes. 

The coordinator tracks the state of the blockchain.
It stores the header of each block and determines the main chain (e.g., with most \pow in Bitcoin and Ethereum).
It instructs the \shards of which blocks to request, process, and validate.
In the case of a \emph{reorg} (rollback of state, which is possible in some blockchains), the coordinator commands the \shards to roll back their state to the block of the divergence and start following the new chain.
\negspace
\paragraph*{Node-Shards}
Each \shard is assigned a local \sid.
The \sid determines which transactions the \shard is responsible for.
Once all \shards are connected, the coordinator sends each \shard the \sid, IP address, and port of all sibling \shards.
Next, \shards connect to form a clique.
Via the \sid, each transaction is mapped to the \shard responsible for storing it and its outputs.

Each \shard stores transactions in a \emph{\textit{\bshard}}~-- a portion of a block that contains only transactions assigned to its \sid. 
It similarly stores UTXOs in a \emph{UTXO-shard}, and pending transactions in its \emph{\mem shard}.

\snegspace
\subsection{System Design}
\label{sub:arch_sesign}

We proceed to describe the details of the components' interaction.
The design sets to achieve several goals that seem contradictory. 
First, transaction lookup should be an~$O(1)$ operation so block validation is quick. 
Transaction to \shard assignment should allow for efficient communication between \shards.
Nevertheless, an adversary should not be able to overload any single \shard. 
Ostraka achieves all these goals as follows. 

Transaction inputs reference previous transaction outputs by \txhash and the \texttt{index} of the output in the transaction.
Each \txhash is unique to a transaction (we use \sha). 
Therefore, \txhash is the natural index for transactions.

\negspace
\paragraph*{Any-order execution}
\label{sec:transaction_order}

In unified blockchain clients, storage and processing are all performed on a single machine (e.g.,~\cite{nakamoto2008bitcoin,wood2014ethereum}), transactions in a block are ordered as a list.
A transaction can only spend outputs created in a previous block or earlier in the same block's list.
We call this common approach, the \emph{topological-order} algorithm.
Using a topological order in \sys would have implied unnecessary dependencies, requiring costly coordination among \shards.
Instead, Ostraka treats each block as an unordered set of transactions.
Dependencies of a transaction are satisfied if the referenced output is in an earlier block or anywhere in the current block.
We justify the correctness of unordered block validation in Section~\S\ref{sec:security}.

\negspace
\paragraph*{Adding salt}
\label{sec:arch_salt}

%
When considering how to distribute transactions equally among \shards, one option to simply take the \txhash of each transaction, and map it to a shard using a deterministic function (e.g. taking the MSBs of the hash).
This method relies on the natural distribution of the \sha value of each transaction.
However, this would allow an attacker to cheaply form transactions that would all be assigned to the same \shard on all nodes.
The processing speed of all nodes would then be~$1/\ell$ (for~$\ell$ \shards), achieving~\dos.
We further analyzed \dos susceptibility in Section~\ref{sub:sec_dos}.

To prevent such attacks, each node generates a random \salt upon initialization.
We use the salt and apply it to each transactions with an additional hash as follows: $ \newtxhash = \sha(\txhash || \salt) $.
By applying a deterministic function to the \newtxhash, we distinctly map a transaction to a \shard, 
Outputs in the \utxoset are stored with the original \txhash, while the \newtxhash is used internally to distribute transaction between \shards.

The coordinator publishes the chosen salt to its \shards and remote nodes.
Thus, peers know the \newtxhash for the specific \salt of the peer, and only sends transactions relevant to the requesting \shard's \sid.
This way network bandwidth is distributed among the \shards and each \shard receives only transactions it requires.
Direct \shard to \shard communication between nodes avoids bottlenecks.

\paragraph{Note} 
Routing all transactions through the coordinator would allow us to keep the salts secret, making DoS attacks harder. 
But as expected, our experiments show that with this design the coordinator becomes a bottleneck. 

\snegspace
        \subsection{Operation}
        \label{sub:arch_operation}

We proceed to describe networking and Ostraka's distributed block validation.

\begin{figure}[t]
    \centering
    \includegraphics[width=0.5\linewidth]{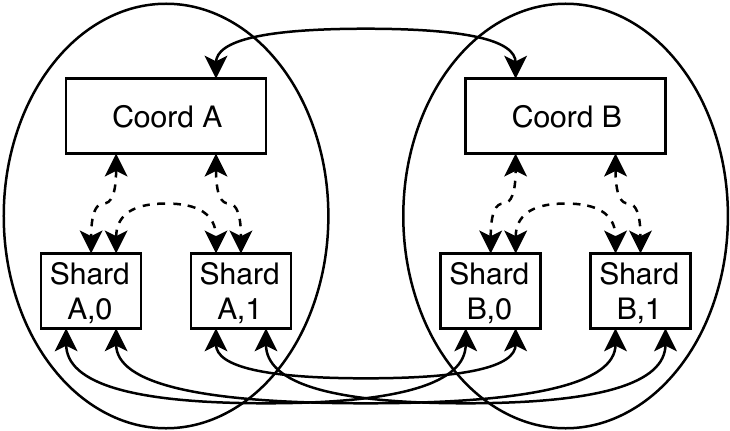}
    \caption{Two connected \sys nodes, with intra-node connections}
    \label{fig:2nodes}
\end{figure}

\paragraph*{Establishing connections}
\label{sub:arch_establish_connection}
Nodes connect to each other according to the system's topology rules (e.g., several random peers in Bitcoin).
When the rules indicate that node~$A$ should connect with node~$B$, the two coordinators connect, and all \shards of node~$A$ connect to all \shards of node $B$.
An example is presented in Fig~\ref{fig:2nodes}

\negspace
\paragraph*{Block validation}
\label{sub:arch_block_validation}
When a node receives a new block, it sends an \inv message to its peers, informing them that a new block has been produced.
The receiving coordinator informs its \shards of the new block hash and the peer who has it.
Each \shard requests the relevant transactions from all the \shards of the peer.
It sends the peer \shards its local \sid and the block hash it is interested in.
Each \shard of the sending node iterates over the transactions associated with the block hash and calculates the \newtxhash using the \salt.
The sending \shard sends the requester only relevant transactions.

Block validation comprises block header validation, where most operations are performed in $\mathcal{O}(1)$, and transaction validation, where operations are $\mathcal{O}(n)$ ($n$ is the number of transactions in the block). 
Block header validation is performed by the coordinator.
It checks for duplicates, \pow, correct reference hash, etc.

The exception is Merkle root validation. The coordinator performs Merkle root validation which is linear in~$n$. 
First, \shards calculate the hashes of their transactions, the leaves of the Merkle tree, and send them to the coordinator.
On a node with~$\ell$ \shards and a block with~$n$ transactions, each \shard is performing~$n / \ell$ hash operations.
The coordinator lexicographically sorts the transactions by hash and then calculates the remaining~$n$ hashes to build the root.
This process could in principle be delegated to \shards by replacing the Merkle tree with an accumulator~\cite{benaloh1993one, baric1997collision}, but their group operations result in performance that is significantly worse than the Merkle tree calculation. 

\shards perform their part of block validation concurrently with the coordinator.
The coordinator awaits approval from all \shards before adding the current block to the chain.
If at any point one of the \shards comes across an illegal transaction, it sends a \bad message to the coordinator, who then discards the block.

Fig.~\ref{fig:protocol} summarizes the steps of block propagation and validation.
To validate a transaction, a \shard must first obtain all the required information.
A \shard only has its share of UTXOs in the \utxoset.
It constructs a list of missing outputs to request from each sibling \shard by using the \txhash, the \salt, and the \sid.
Requests to each \shard are sent in a batch.
Upon request, each \shard sends the transaction output to the requester.

Once an output has been requested, it is marked as \textit{spent}.
Multiple requests for the same output indicate a double-spend attempt, and will thus trigger a \bad message to the coordinator.
Once it receives all transaction outputs, the \shard proceeds to validate the scripts.
These checks include:

\begin{itemize}
    \item{Equal sum}: For each transaction, the aggregate amount of inputs is at least as high as the outputs.

\item{No double spends}: Each input references an output that was produced but not spent.

\item{Authorization}: Each input complies with the spending condition of the output it consumes.
\end{itemize}

\negspace
\paragraph*{Other operations}
Other inter-node operations are easily implemented in \sys.
Peers can request transaction directly from \shards, and block headers from the coordinator.

One special case is that of a reorg.
A reorg happens when a node needs to switch the chain it follows.
The coordinator and \shards must roll back to a state of a previous block, then update it to the longest chain.

A reorg is performed as follows:
First, the coordinator sends the hashes of blocks to roll back, as well as the hashes of the new blocks.
Each \shard fetches the spent outputs records of the rolled-back block from the database.
Each \shard adds the spent outputs of each transaction in the block to the UTXO-shard.
Next, it removes all outputs created by transactions in the blocks-shard from the UTXO-shard.
No communication is required among the \shards during the rollback, and it can thus be quickly resolved.

\begin{center}
\tikzset{
  every node/.append style={font=\large},
  every picture/.append style={
  transform shape,
          scale=0.5
    }
}
\begin{figure}
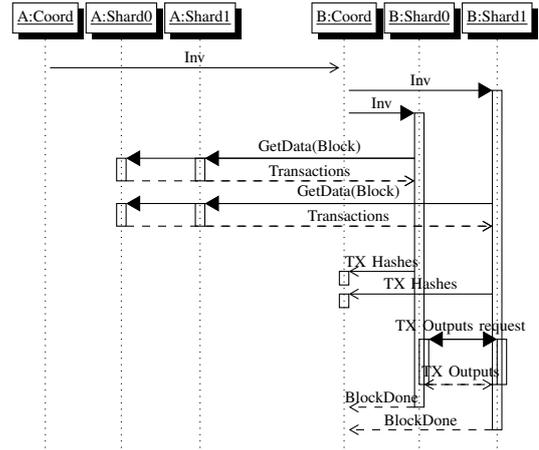

  \centering
  \begin{sequencediagram}
    \newinst{C0}{A:Coord}{}
    \newinst{C00}{A:Shard0}{}
    \newinst{C01}{A:Shard1}{}
    \newinst[2]{C1}{B:Coord}{}
    \newinst{C10}{B:Shard0}{}
    \newinst{C11}{B:Shard1}{}
    \mess[0]{C0}{Inv}{C1}
    \begin{call}{C1}{Inv}{C11}{BlockDone}
        \begin{call}{C1}{Inv}{C10}{BlockDone}
            \postlevel
            \begin{call}{C10}{GetData(Block)}{C01}{Transactions}
            \end{call}
            \prelevel
            \prelevel
            \begin{call}{C10}{}{C00}{}
            \end{call}
            \begin{call}{C11}{GetData(Block)}{C01}{Transactions}
            \end{call}
            \prelevel
            \prelevel
            \begin{call}{C11}{}{C00}{}
            \end{call}
            \postlevel
            \begin{messcall}{C10}{TX Hashes}{C1}{}
            \end{messcall}
            \prelevel
            \begin{messcall}{C11}{TX Hashes}{C1}{}
            \end{messcall}
            \setthreadbias{east}
            \begin{call}{C10}{TX Outputs request}{C11}{TX Outputs}
            \prelevel
            \begin{call}{C11}{}{C10}{}
            \postlevel
            \end{call}
            \prelevel
            \end{call}
            \setthreadbias{center}
        \end{call}
    \end{call}
  \end{sequencediagram}
    \caption{\sys block transmission and validation messages.}
    \label{fig:protocol}
\end{figure}
\end{center}

\newcommand{\sht}{\ensuremath{t_{\textit{hop}}}\xspace}
\newcommand{\bpt}{\ensuremath{Bpt}\xspace}
\newcommand{\btt}{\ensuremath{Btt}\xspace}

\bnegspace
\section{Security}
\label{sec:security}

\sys can be used in conjunction with many consensus protocols (e.g.,~\cite{yin2019hotstuff,ng2016eyal,byzcoin2016kogias,ouroboros2017kiayias,rocket2018snowflake}). 
We thus need to prove that a system using \sys is at least as secure as a system using a unified node.

We start by showing that Ostraka behaves as a unified node~(\S\ref{sec:sec_functional}).
Beyond functionality, blockchain protocols rely on propagation time to maintain security.
Therefore we also show that Ostraka's design does not enable attacks on node processing time.
We show how individual-node shuffling prevents an adversary from affecting macro system performance~(\S\ref{sub:sec_dos}).

\newcommand{\accept}{\texttt{accept}\xspace}
\newcommand{\reject}{\texttt{reject}\xspace}
\newcommand{\invalid}{\texttt{invalid}\xspace}
\newcommand{\rejects}{\textsc{rejects}\xspace}
\newcommand{\spent}{\textit{spent}\xspace}

\newcommand{\utxost}{\ensuremath{\texttt{UTXO\_set}}\xspace}
\newcommand{\utxosti}[1]{\ensuremath{\texttt{UTXO\_set}_{#1}\xspace}}
\newcommand{\utxosd}{\ensuremath{\texttt{UTXO\_shard}}\xspace}

\newcommand{\algT}{\ensuremath{\mathcal{A}_T}\xspace}
\newcommand{\algL}{\ensuremath{\mathcal{A}_U}\xspace}
\newcommand{\algO}{\ensuremath{\mathcal{A}_O}\xspace}

    \subsection{Functional Indistinguishability}
    \label{sec:sec_functional}

Blockchain systems typically~\cite{nakamoto2008bitcoin,wood2014ethereum,byzcoin2016kogias,ng2016eyal} form blocks that are ordered such that each transaction follows transactions it depends on.
We call this a \emph{topological-order System}.

In contrast, Ostraka is agnostic the internal order. 
We call such a system an \emph{unordered system}.
Of course, topologically ordered blocks may be invalid in a system that expects lexicographically ordered blocks, and vice versa.
The \sys algorithm would except a block created by Bitcoin miners, but not every out-of-order block,  would be accepted in Bitcoin without changes to the validation rules.
We thus need to show that any \emph{set} of transactions can be processed as a legal block in a topological system if and only if they can be processed as a legal block in the unordered system.
We leave it for future work to explore if \sys can be integrated with Bitcoin as it exists today.
As the resulting Merkle root depends on transaction ordering, this might not be possible without a hard fork.

We show that an unordered system is functionally indistinguishable from a Topological-order system.
\snegspace
\paragraph*{From topological to unordered systems}

Most stages of block validation (cf.~\S\ref{sub:arch_block_validation}) like signature and \pow verification are agnostic to transaction order.
Two exceptions are the Merkle tree calculation and transaction dependency checks.

Merkle tree calculation depends on the transaction order, but the choice of the order does not matter as long as all nodes calculate it consistently.
For this purpose, Ostraka orders transactions lexicographically, according to their \txhash.

It remains to show that any \emph{set} of transactions can be processed as a legal block in a topological system if and only if they can be processed as a legal block in the unordered system.

To formally state this claim we first specify the two block-validation algorithms.
Both take a UTXO set and a set of transactions and return either the updated UTXO set or~$\bot$ if the transactions cannot form a legal block.

The topological-order system algorithm denoted~\algT, is shown in Algorithm~\ref{alg:syst}.
The algorithm receives a~\utxost and the set of transactions~$T$.
The algorithm first sorts~$T$ topologically and then processes the transactions in order.
It is important to note that the set~$T$ can always be ordered topologically: cycles between transactions cannot form since the references are by transaction hash.
Assuming~\sha can be modeled as a random oracle, Harris~\cite{harris1960probability} shows that for a space of~$2^\kappa$, the expected number of steps to find a cycle is $\frac{1}{4}\sqrt{2\pi 2^\kappa}$, therefore an adversary cannot find such a cycle except with negligible probability in the hash output size.
Therefore the sorting algorithm is well defined.

The algorithm returns \empt if at some point it encounters a transaction whose input is in not in the~\utxost.
For each transaction, the operation \textit{Valitates TX} (line~\ref{alg:syst:validate}), covers all additional transaction validation steps after the referenced outputs were found, e.g., script validation, correct sum etc.
If \textit{Validate TX} fails, it returns \empt.

The unordered system algorithm denoted~\algL, is shown in Algorithm~\ref{alg:sysl}.
Unlike~Algorithm \algT, in~\algL transactions are not ordered.
All outputs are first added to the~\utxost, then the algorithm processes the transactions sequentially in an arbitrary order.
Other than that, the operation is similar to~\algT.

\begin{algorithm}[t]
\label{alg:syst}
\DontPrintSemicolon
\SetAlgoLined
\SetKwInOut{Input}{Input}\SetKwInOut{Output}{Output}
\Input{Set of transactions $T$, \utxost}
\Output{New \utxost or \empt}
\BlankLine

$(t_1, t_2, \dots, t_k) \gets T$ sorted in topological order \label{alg:sort}
\BlankLine
\ForEach{$\textit{TX} \gets t_1, t_2, \dots, t_k$}{
    \ForEach{\inps in \textit{TX}}{
        \uIf{$\inps \not \in \utxost$} {
            return \empt \label{syst:bot}
        }
        \uElse{
            $\utxost \gets \utxost \setminus \{\inps\} $
        }
    }
    \ForEach{\txo in \textit{TX}}{
        $\utxost \gets \utxost \cup \{ \txo \}  $
    }
    \textit{Validate TX} \label{alg:syst:validate}
}
return \utxost
\caption{Topological \algT}
\end{algorithm}

\begin{algorithm}
\label{alg:sysl}
\DontPrintSemicolon
\SetAlgoLined
\SetKwInOut{Input}{Input}\SetKwInOut{Output}{Output}
\Input{Set of transactions $T$, \utxost}
\Output{New \utxost or \empt}
\BlankLine

\ForEach{$\textit{TX} \in T$} {
    \ForEach{\txo in \textit{TX}}{
        $\utxost \gets \utxost \cup \{\txo\}$ \label{sysl:add}
    }
}
\BlankLine
\ForEach{$\textit{TX} \in T$} {
    \ForEach{\inps by \textit{TX}}{
        \uIf{$\inps \not \in \texttt{\utxost}$} {
            return \empt \label{sysl:bot}
        }
        \uElse{
            $\utxost \gets \utxost \setminus \{\inps\}$
        }
    }
    \textit{Validate TX}
}
return \utxost
\caption{Unordered (\algL)}
\end{algorithm}

We show that Algorithms~\algT and~\algL are functionally indistinguishable.

\newcommand{\lemAlgLAlgT}{
Given the same \utxost and set of transactions $T$, \algT and \algL return the same value.
}
\begin{lemma} \label{lem:alglalgt}
\lemAlgLAlgT
\end{lemma}

We prove the lemma by showing that~\algT returns~\empt if and only if~\algL returns~\empt: 

\snegspace
\begin{proof}
We prove in both directions.

\paragraph*{$\algT \to \algL$}
We first show that if \algT rejects an input tuple~$(T, \utxost)$ then \algL also rejects this input.

If \algT returns~\empt, there exists some transaction $\tx{i} \in T$, at index $i$, and an input in \tx{i}, referencing an UTXO $\inps \not \in \utxost$, produced by an output \txo.
This can be a result of three cases, and we show that \algL rejects as well.

\begin{enumerate}[leftmargin=*, wide=0pt, align=left]
    \item The UTXO is not in the \utxost and is not produced by any transaction preceding \tx{i}:
There can be no transaction succeeding \tx{i} which produces \txo as it contradicts the topological order, thus, there is not \emph{any} transaction producing output \txo in $T$.
Upon processing \tx{i}, \algL return \empt(Algorithm~\ref{alg:sysl} line~\ref{sysl:bot}).

    \item  The UTXO is not in the \utxost and is produced by some transaction \tx{k}, preceding \tx{i} but consumed by some transaction \tx{j}, succeeding \tx{k} but preceding \tx{i}:
        In \algL, \txo is added to the \utxost in line~\ref{sysl:add}, \algL processes either \tx{j} or \tx{i} first.
    Without loss of generality, \tx{i} is processed first.
     Upon processing \tx{j}, \algL would not find UTXO \inps in the \utxost and return \empt (Algorithm~\ref{alg:sysl} line~\ref{sysl:bot}).

    \item The UTXO is in the \utxost and it is consumed by some transaction \tx{j}, preceding \tx{i}:
Similarly to the previous case, except now the UTXO is in the \utxost.

\end{enumerate}

\paragraph*{$\algL \to \algT$}
If \algL returns \empt, there exists some transaction $\tx{i} \in T$, and an input \inps in \tx{i}, referencing an UTXO $\inps \not \in \utxost$ produced by an output \txo.
This can also be a result of three cases, and we show that \algT rejects as well.

\begin{enumerate}[leftmargin=*, wide=0pt, align=left]
    \item The UTXO is not in the \utxost and is not produced by \emph{any} transaction in $T$:
    In \algT, the UTXO is not produced by \emph{any} transaction in $T$ and specifically, by any transaction preceding \tx{i} in the topological order of $T$ in \algT.
Thus \algT return \empt .

    \item  The UTXO is not in the \utxost and is produced by some transaction \tx{k}, but is also consumed by some transaction \tx{j} :
In the topological order of $T$, both \tx{j} and \tx{i} must succeed \tx{k}.
Transaction \tx{k} is processed first and adds \inps to the \utxost.
Without loss of generality, \tx{i} is processed first and removes \inps from the \utxost.
Upon processing \tx{j}, output \inps is not in the \utxost and system $T$ return \empt (Algorithm~\ref{alg:syst} line~\ref{syst:bot}).

    \item The output is in the \utxost and it is also consumed by some transaction \tx{j}:
    Similar to the second case, but \inps is in the \utxost.

\end{enumerate}

Therefore, given the same \utxost, and transactions $T$, \algT return \empt if and only if \algL returns~\empt.
Both algorithms add the same set of \textit{outputs} to the \utxost, and remove the same set of \textit{utxos}.
No output can be added to the \utxost twice, and any attempt to remove an output twice results in \empt.
Thus the new \utxost returned by each algorithm is identical.
\end{proof}

\negspace
\paragraph*{From unified node to distributed node}
\label{sub:sec_uni_to_dist}

We now show that \sys is functionally indistinguishable from~\algL.
The distributed algorithm of Ostraka, denoted~\algO, is shown in Algorithms~\ref{alg:syso_coord}--\ref{alg:syso_shard}.
A Coordinator algorithm receives the \utxost and transaction set $T$ as input.
The Coordinator algorithm first distributes the \utxost and transactions among~$\ell$ \shards.
\shards begin validation by adding all produced outputs in their set of transactions $T_i$, to their local \utxosd.
They proceed by requesting missing output information from the relevant siblings.
Once they receive the information, each \shard proceeds to validate its set of transactions.
If some validation stage fails the \shard sends \empt, otherwise the \shard sends its updated \utxosd.
If all \shards returned updated sets, the coordinator returns their union, otherwise \empt.
The pseudo-code makes a slight simplification of Ostraka that nonetheless represents its logical behavior.

\newcommand{\lemAlgLAlgO}{
Given the same \utxost and set of transactions $T$, \algL and \algO return the same value.
}

\begin{algorithm}[t]
\label{alg:syso_coord}
\DontPrintSemicolon
\SetAlgoLined
\SetKwInOut{Input}{Input}\SetKwInOut{Output}{Output}
\SetKwFor{ForAll}{for all}{do in parallel}{end}
\SetKwFor{Upon}{upon}{do}{}
\Input{Set of transactions $T$, \utxost}
\Output{New \utxost or \empt}
\BlankLine
Distribute $T$ into $T_1 \dots T_\ell$ and \utxost into \utxosti{1} \dots \utxosti{\ell}\;
\ForAll{\shards, $S_i \gets S_1 \dots S_\ell$}{
    Send ($T_i$, \utxosti{i}) to \shard $S_i$
}
\caption{\sys, \algO}
\For{\shards, $S_i \gets S_1 \dots S_\ell$} {
    Receive \utxosti{i}\;
    \uIf{Received \empt} {
        Return \empt
    } \uElse {
    $\utxost \gets \utxost \cup \{\utxosti{i}\}$
   }

}
Return \utxost
\end{algorithm}

\begin{algorithm}[t]
\label{alg:syso_shard}
\DontPrintSemicolon
\SetAlgoLined
\SetKwInOut{Input}{Input}\SetKwInOut{Output}{Output}
\SetKwFor{ForAll}{for all}{do in parallel}{end}
\SetKwFor{Upon}{upon}{do}{}
\Input{Set of transactions $T_i$, \utxosd}
\Output{New \utxost or \empt}
\BlankLine

Add all produced outputs in $T_i$ to the \utxosd \label{s:add}\;

\tcc{Request TX outputs}

Request all referenced outputs in $T_i$ from siblings\;
Receive requests for missing outputs\;

\ForEach{\inpr}{
    \uIf{$\inpr \not \in \utxosd$}{
        Send \empt to Coordinator \label{s:remote}
    }
    \uElse{
        $\utxosd \gets \utxosd \setminus \{\inpr\} $
    }
}

\tcc{Receive TX outputs}

Send requested outputs to requesters\;
Receive requested outputs and add to \utxosd \;

\BlankLine

\tcc{Transaction validation}
\ForEach{$\textit{TX} \gets t_1, t_2, \dots, t_k \in T_i$}{
    \ForEach{\inps by \textit{TX}}{
        \uIf{$\inps \not \in \utxost$} {
            Send \empt to Coordinator \label{s:local}
        }
        \uElse{
            $\utxosd \gets \utxosd \setminus \{\inps\} $
        }
    }
    \textit{Validate TX}
}
return \utxosd
\caption{\sys shard algorithm}
\end{algorithm}

We show that an unordered-system algorithm and Ostraka are functionally indistinguishable.

\begin{lemma} \label{lem:alglalgo}
\lemAlgLAlgO
\end{lemma}

As before, we prove the lemma by showing that~\algL returns~\empt if and only if~\algO returns~\empt: 
\snegspace
\begin{proof}
We prove in both directions.

\negspace
\paragraph*{$\algL \to \algO$}
We first show that if \algL rejects an input~$(T, \utxost)$ then \algO also rejects this input.

If \algL returns~\empt, there exists some transaction $\tx{i} \in T$, and an input  in \tx{i}, referencing an UTXO $\inps \not \in \utxost$, produced by an output \txo.

\begin{enumerate}[leftmargin=*, wide=0pt, align=left]
    \item The UTXO is not in the \utxost and is not produced by \emph{any} transaction in $T$: The UTXO is not produced by any transaction in $T$, thus no \shard receives a transaction producing \inps in ~\algO.
WLOG, some \shard $S_i$ receives \tx{i}.
If \inps references an output assigned to a \shard $S_j$, when receiving request for \inps, $S_j$ does not have \inps in its \utxosd and send \empt (Alg.~\ref{alg:syso_shard} line~\ref{s:remote}).
If \inps references a transaction assigned to $S_i$, when processing transactions, \inps is not in \utxosd of $S_i$, and $S_i$ sends~\empt (line~\ref{s:local}). 

    \item  The UTXO is not in the \utxost and is produced by some transaction \tx{k}, but is also consumed by some transaction \tx{j}:
WLOG, \txo is produced by a transactions assigned to some \shard $S_k$, and is added to its \utxosd (Algorithm~\ref{alg:syso_shard} line~\ref{s:add}).
If both consuming transactions \tx{j} and \tx{i}, are assigned to sibling \shards, then upon receiving all requests from sibling \shards, $S_k$ processes one of the requests first and remove \inps from it \utxosd.
Upon processing the second request, \inps is not in \utxosd, and $S_k$ sends \empt (Algorithm~\ref{alg:syso_shard} line~\ref{s:remote}).
If \tx{j} is assigned to a sibling \shard, and \tx{i} is assigned to $S_k$, \inps is removed from \utxosd after all outputs are requested.
Upon processing \tx{i}, $S_k$ does not have \inps in \utxosd, and send \empt (Algorithm~\ref{alg:syso_shard} line~\ref{s:local}).
If both \tx{i} and \tx{j} are assigned to $S_k$, one of them is be processed first and removes \inps from the \utxosd, and upon processing the second one, \inps is not be in \utxosd and $S_k$ sends \empt (Algorithm~\ref{alg:syso_shard} line~\ref{s:local}).

    \item The UTXO is in the \utxost and it is also consumed by some transaction \tx{j}: This is similar to the second case, but \txo is in the \utxosd of $S_k$.

\end{enumerate}

\negspace
\paragraph*{$\algO \to \algL$}

    If \algO returns \empt and rejects, then some \shard $S_i$ sends \empt to the Coordinator.
    This can be a result of several cases, and we show that \algL rejects as well.

First, notice that the \utxost and transactions $T$ are distributed deterministically.
There can be no two \shards producing the same \txo, and similarly, \inps cannot reside in two UTXO-shards.

\begin{enumerate}[leftmargin=*, wide=0pt, align=left]
    \item When processing requests for outputs, there is some UTXO $\inps \not \in \utxosd$
    \begin{enumerate}
    \item The UTXO is not in the \utxosd, is not produced by any transaction in $T_i$, and is requested by some sibling \shard $S_j$. Then there exists some transaction \tx{j} $ \in T$ consuming UTXO. As \utxost is distributed deterministically, there can be no other \utxosd other than \utxosti{i} where the UTXO might reside. Therefore, it is not in the \utxost of \algL, and upon processing \tx{j}, \algL returns \empt (Algorithm~\ref{alg:sysl} line~\ref{sysl:bot}).

    \item The UTXO is not in the \utxosd, is produced by some transactions \tx{k}, and is requested by some two sibling \shard $S_j$, $S_l$. Then there exist some two transaction \tx{j} and \tx{j} $\in T$, consuming the same UTXO. WLOG, in \algL, \tx{j} is processed first and consumes the UTXO. Upon processing \tx{l}, \algL returns \empt (Algorithm~\ref{alg:sysl} line~\ref{sysl:bot}).
    \end{enumerate}

\item When processing transactions, there exists some transaction $\tx{i} \in T_i$, and an input in \tx{i}, referencing an UTXO $\inps \not \in \utxosd$.
    \begin{enumerate}
        \item The UTXO is in the \utxosd and is requested by some sibling \shard $S_j$:
        There is some other transaction \tx{j} $\in T$ consuming the UTXO. WLOG, in \algL \tx{j} is processed first, and upon processing \tx{i}, \algL returns \empt (Algorithm~\ref{alg:sysl} line~\ref{sysl:bot}).

        \item The UTXO is not in the \utxosd, is produced by some transaction \tx{k}, and is requested by some sibling \shard $S_j$. The case is similar to the previous after all outputs are added to the \utxost in Algorithm~\algL (Algorithm~\ref{alg:sysl} line~\ref{sysl:add}).

        \item The cases enumerated for \algL are the same if all transactions are assigned to a single \shard, and are not requested by any sibling.
    \end{enumerate}

\end{enumerate}

Therefore, given the same \utxost, and transactions $T$, \algL return \empt if and only if \algO returns~\empt.
\end{proof}

We can now show that a classical topological-order system with a unified node is functionally indistinguishable from an Ostraka node.
\begin{theorem}
Given the same \utxost and set of transactions $T$, a topologically ordered system and \sys, are functionally indistinguishable.
\end{theorem}

The proof follows directly from Lemmas~\ref{lem:alglalgt} and~\ref{lem:alglalgo}.

\subsection{Processing Time}
\label{sub:sec_dos}

We have shown the systems are functionally indistinguishable; it remains to show that an attacker cannot affect block processing time in \sys more than it could in a unified node.
Limiting processing time is important to prevent an attacker from strangling the system, creating a major slowdown and loss of throughput.

\negspace
\paragraph*{Targeting a single \shard}
In a unified node, block processing time is (roughly) the sum of processing all transactions.
In Ostraka it is the time it takes the slowest \shard to process its transactions.

If transaction distribution is identical for all nodes, then an attacker can cheaply generate transactions that are all placed in the same \shard.
For example, an attacker can create a block where all transactions start with multiple zeros in the MSB.
In general, to create a transaction starting with $\ell$ zeros, the attacker needs $2^\ell$ attempts in expectation.
For a block of $L$ transactions, a total of $L \cdot 2^\ell$ attempts are required in expectation.
Note that this analysis applies to any index function.

We denote by \msb the distribution function applied to \txhash to determine the \shard responsible for it.
For $\ell$ \shards, the \msb results in a number from $1$ to~$\ell$.

In \sys, we use a \salt to redistributed transactions.
Each node is aware of the salt values of its neighbors.
As connections are performed randomly, it is unlikely a node will obtain the values of all nodes in the system.
Yet, even if one obtains all \salts, attack difficulty rises exponentially with the number of targeted nodes.

Assume the attacker knows the \salt values of $ k $ peers, $ S_1..S_k $.
We define the distribution function \msb for a value $x$ and a \salt value $S_i$ to be ${\msb(x, S_i)~=~\sha(x || S_i)}$.
The first transaction the attacker creates will result in some value $v_i$ for each of the $k$ peers.
The attacker now needs every subsequent transaction to result in the same value $v_i$  for each peer.
The attacker thus needs to find a \txhash such that:
${ \forall 1 \le i \le k:  \msb(\txhash, S_i)~=~v_i}$.
The probability of finding such value, if the attacker tries to target one of $\ell$ \shards is $(1/\ell)^{k}$.
To cause an $\ell$  times slowdown for $k$ peers, requires ${\ell}^k$ attempts per transaction in expectation.

It quickly becomes infeasible to affect even a small percentage of the system.
For example, creating one transaction affecting the validation time of 100 nodes requires the hashing rate of Bitcoin for 800,000 years~\cite{btcHashRate}.

\negspace
\paragraph*{Creating a single large transaction}
Creating an abnormally large single transaction occupying an entire block can also cause a significant slowdown.
In \sys, each transaction is processed by a single \shard.
A huge transaction that occupies an entire block, would annul the benefits of the distributed architecture.
Adding \salt does not help mitigate this attack vector.
In \sys, we mitigate this attack by adding a transaction size limit.
The limit should be proportional to the block size, to allow even distribution of transactions across multiple \shards.
In general, a user can make any payment with only two types of transactions.
A \emph{join} transaction of two outputs into one, or a \emph{fork} transaction of a single output into two.

\section{Evaluation}

\label{sec:evaluation}
\begin{figure*}[!htb]
\begin{minipage}[b]{.32\textwidth}
    \includegraphics[width=1.0\linewidth]{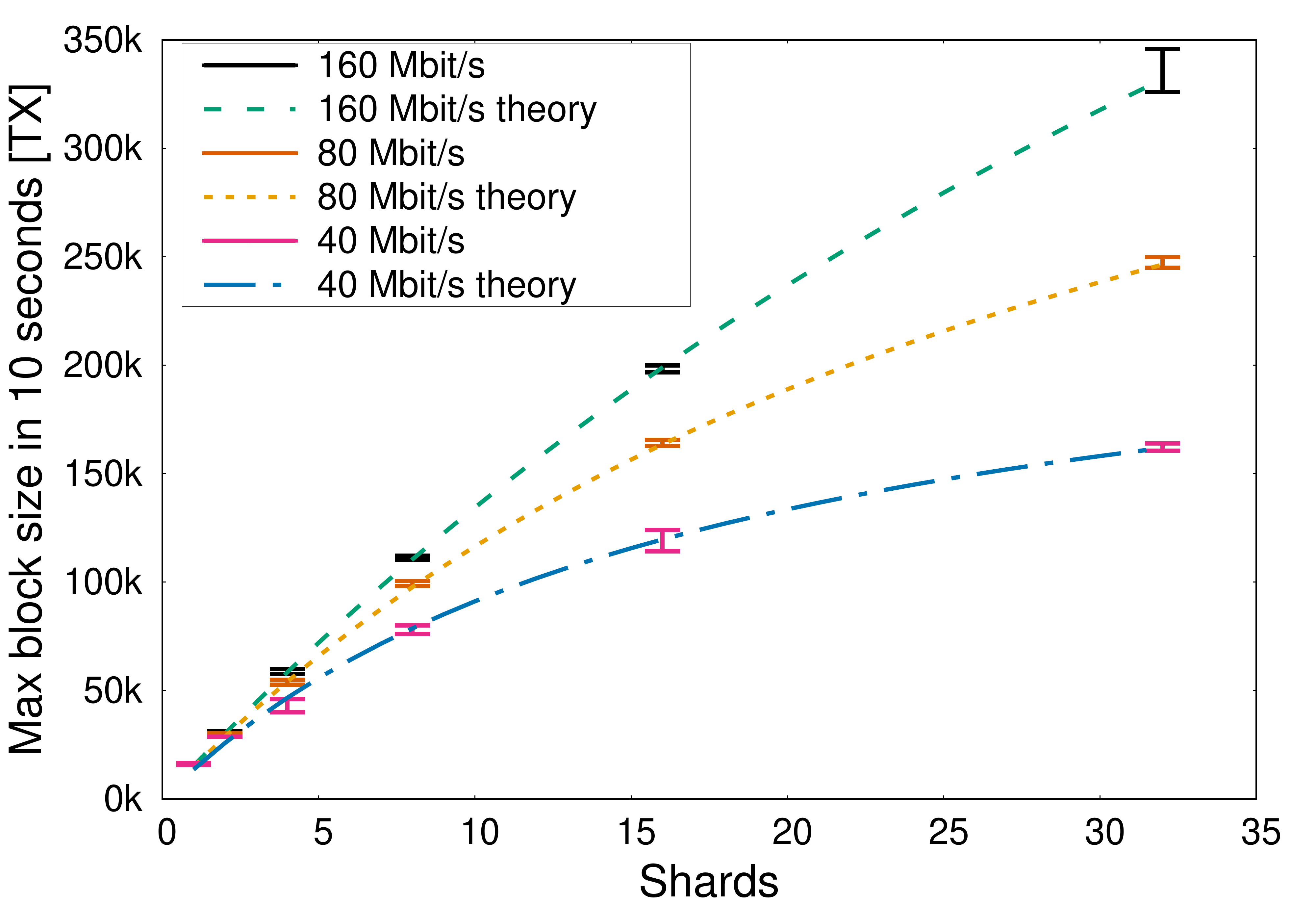}
    \caption{Theoretical and measured capacity, with variable available network bandwidth}
    \label{fig:cap_estimate}
\end{minipage}
\hfill
\begin{minipage}[b]{.32\textwidth}
  \centering
    \includegraphics[width=1.0\linewidth]{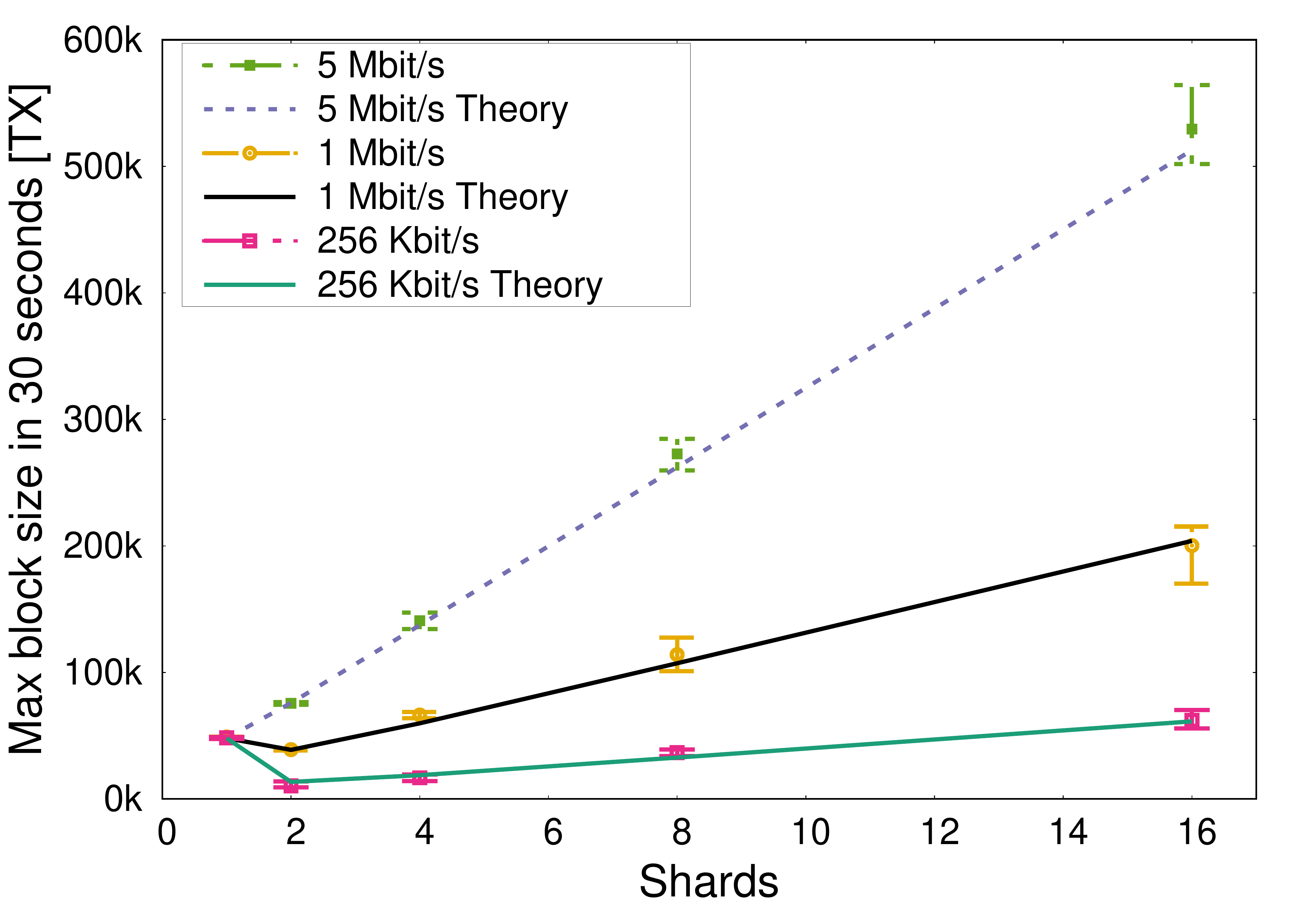}
    \caption{Theoretical and measured capacity with limited intra-node bandwidth}
    \label{fig:intra_shard}
\end{minipage}%
\hfill
\begin{minipage}[b]{.32\textwidth}
    \includegraphics[width=1.0\linewidth]{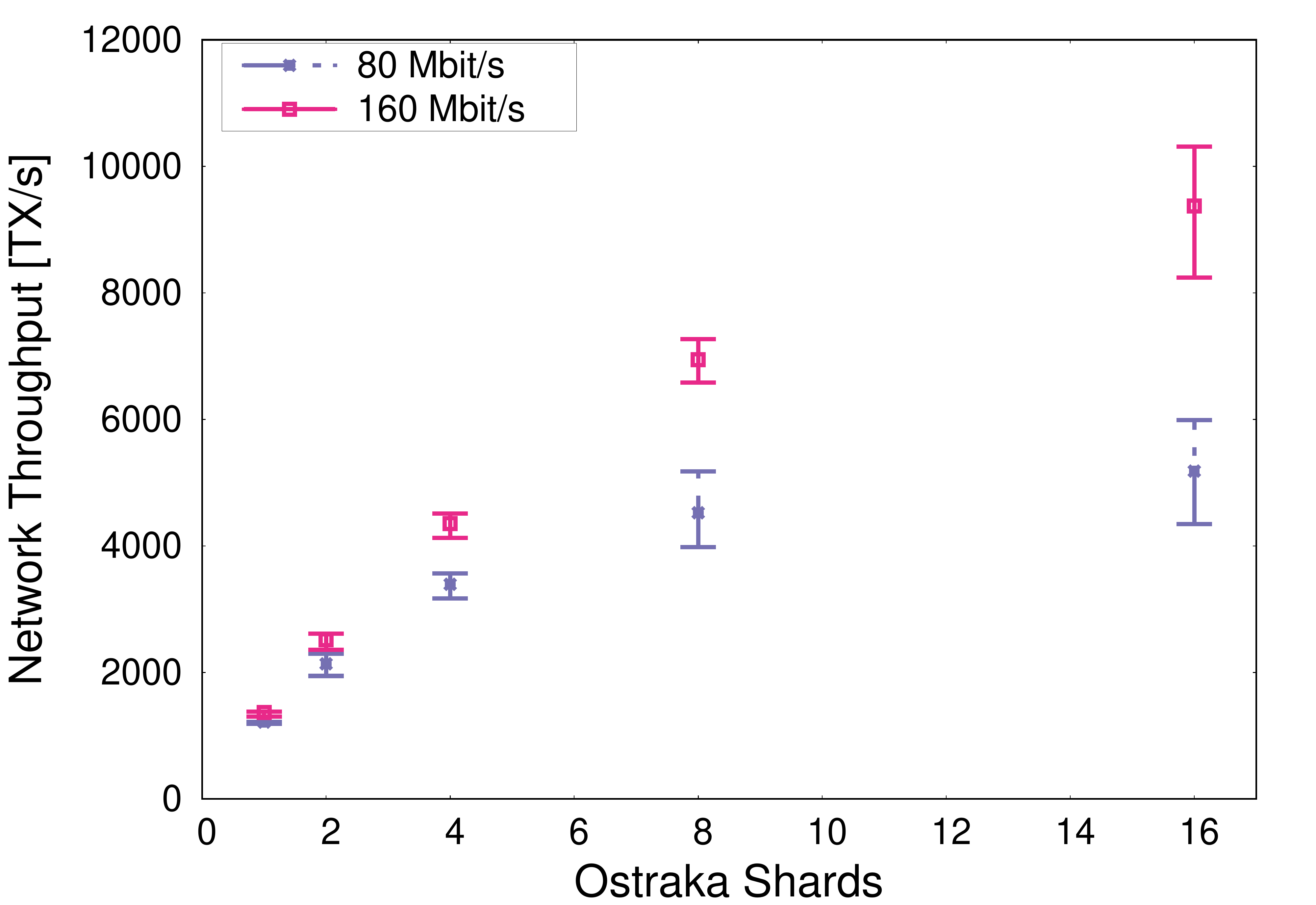}
    \caption{Bitcoin-NG simulation with varying bandwidth and \shards per node}
    \label{fig:bitcoin_ng}
\end{minipage}
\end{figure*}

The throughput of a blockchain depends on \emph{\sht}:~the time it takes for a single node to receive and process the block.
\sht depends on both block transmission time \btt and block processing time \bpt.
Each node validates the block before it propagates it further, to prevent \dos attacks on the system~\cite{croman2016scaling,decker2013information}.

We present the measurements and theoretical analysis of \sht.
We begin by describing our evaluation methodology (\S\ref{sub:eval_methodology}), and proceed to measure how the number of \shards and the node bandwidth affects the \sht~(\S\ref{sub:eval_processing}).
Similarly, we analyze and present the effect of intra-node bandwidth~(\S\ref{sub:eval_intra_shard}). We conclude with a network simulation~(\S\ref{sub:eval_simulation}).

\newcommand{\BS}{\ensuremath{\textit{BSize}}\xspace}
\newcommand{\MB}{\ensuremath{\textit{MBytes}}\xspace}
\newcommand{\TXsize}{\ensuremath{\textit{TxSize}}\xspace}
\newcommand{\Nshards}{\ensuremath{\textit{\#\shards}}\xspace}
\newcommand{\Data}{\ensuremath{\textit{TX}_{out}}\xspace}
\newcommand{\BW}{\ensuremath{\textit{BW}}\xspace}
\newcommand{\Sec}{\ensuremath{\textit{sec} }\xspace}
\newcommand{\Cap}{\ensuremath{\shard_{\textit{tps}}}\xspace}
\newcommand{\BWS}{\ensuremath{{\textit{BW}_\shard}}\xspace}
\newcommand{\tv}{\ensuremath{t_v}\xspace}

\subsection{Methodology}
\label{sub:eval_methodology}
We have implemented the necessary elements for performance evaluation based on the existing \emph{btcd} Bitcoin client~\cite{btcd}.
We implement the intra-node and inter-node networking, as well as our distributed block validation protocol.
In total, we have added ~10K LoC in Go and an additional 2K LoC of Python scripts for testing and simulation.
Our implementation is available as open source~\cite{ostrakaimpl}. 

We measure block sending time between two \sys nodes and block processing time on the receiving node.
We used a \texttt{c4.large} EC2 instance (2 vCPUs 4GB RAM) for each \shard and a \texttt{c4.xlarge} instance (4 vCPUs 8GB RAM) for the coordinator of each node.
A block comprises transactions with two inputs and two outputs, i.e., about~250 \textit{Bytes}, which is the median Bitcoin transaction size.
Each single EC2 \shard can process approximately $ 1700\  \textit{Tx}/\textit{sec}$.
Each block undergoes full validation, i.e. we do not assume some of the transactions have been previously known to the receiving node, and have already been validated.
Transaction dependency within the block does not affect processing time, as transaction validation is performed in distributed, achievable by our block validation algorithm.
Each figure with error bars presents the minimum, maximum, and average measurements.
Theoretical analysis is presented where applicable.

\subsection{Block time and node capacity}
\label{sub:eval_processing}
Total block time \sht depends on both block transmission time \btt and block processing time \bpt.
We begin by isolating and evaluating the \bpt.
We measure \bpt of various configurations, without restriction on intra-node or inter-node bandwidth.

The number of transactions in a block is the block size, denoted~\BS, divided by the mean transaction size, denoted~\TXsize.
\Cap is the number of transactions per second a \shard can process.
\bpt is thus the number of transactions in a block divided by all \shards combined processing rate : $\bpt = {\frac{\BS}{\TXsize} \cdot \frac{1}{\Cap \cdot \Nshards}}$.

We measure the \bpt with an increasing number of transactions.
We repeat this experiment varying the number of \shards from~1 to~32.
We estimate the number of transactions each configuration processes in ten seconds from a linear regression.
We find the maximum block size a node is capable of processing within the limit.
Ten seconds was chosen arbitrarily to be an upper limit on the processing time of each node.
The two \textit{Unlimited} plots of Figure~\ref{fig:cap_estimate} show our measurements.
The theoretical unlimited bandwidth line represents the best achievable results, multiplying the processing capacity of a single \shard by a factor.
The error bars represent our actual measurements. 
We observe that \sys achieves near-optimal improvement as we increase the number of \shards.

Next, we add the effect of block transmission time (\btt) on \sht.
\btt is determined by network latency and bandwidth.
The bandwidth, denoted \BW, is the total network bandwidth available to all the \shards combined.
We focus on network bandwidth as it is more significant than the latency for large blocks~\cite{croman2016scaling}.
We define a node's \emph{capacity} as the number of total transactions per second a node can receive and processes.
The capacity is thus the total number of transactions in a block $\frac{\BS}{\TXsize}$, divided by \sht.

\begin{equation}
\textit{Capacity} \left[\frac{\textit{TX}}{\textit{sec}} \right] =
\frac{1}{\frac{\TXsize}{\BW} +
\left(\frac{1}{\Cap\cdot \Nshards}\right)}
\end{equation}

By adding more \shards we expect \sht to approach \btt, as \bpt approaches 0.
We measure capacity when both the network bandwidth and the number of \shards vary.
We measure with configurations of~$\textit{1,2,4,8,16}$ and $32$ \shards,
as well as total bandwidth of $40$ and $160\textit{Mbit/s}$ for the combined node.
As before, we estimate capacity by evaluating the linear regression at ten seconds.

Fig~\ref{fig:cap_estimate} shows capacity measurements as well as theoretical analysis.
Error bars represent our measurements for each configuration of bandwidth and \shards.
The dashed lines represent our theoretical analysis as shown in equation (1).
Our measurements match our theoretical evaluation: as we increase the number of \shards, the capacity of the node increases, yet limited by the available bandwidth.

        \subsection{Intra-node network}
        \label{sub:eval_intra_shard}

Part of the \sys protocol is requesting and sending outputs between \shards.
When intra-node bandwidth is high, time to send requests and later to send information to requesting \shards is not significant.
Ostraka can be implemented to run on a single system, where \shards communicate over Unix sockets, with much higher bandwidth.
However, we wanted to also examine the case of several distant \shards with limited bandwidth between them.
Limiting the intra-node bandwidth would affect the \bpt.

For $n$ transactions per block and $\ell$ \shards, each \shard has in expectation $\frac{n}{\ell}\frac{\ell-1}{\ell}$ outputs it needs to request from its siblings.
This is also the number of outputs it is required to send.

It is important to note that although the number of siblings each \shard must contact increases with the number of \shards, the overall number of outputs requested/sent per \shard decreases.
Thus, intra-node communication benefits and decreases linearly as we add additional \shards.

We denote by \BWS the intra-node bandwidth and by \Data, the expected size required for sending and receiving outputs information per transaction.
\bpt, when accounting for intra-node communication is thus:
$$
    \bpt = \frac{\BS}{\TXsize}\cdot\frac{1}{\Nshards}\cdot\left(\frac{1}{\Cap} +
        \frac{\Nshards - 1}{\Nshards} \cdot \frac{\Data}{\BWS}
    \right)
$$

We measure \bpt while restraining intra-node bandwidth to~$5 \textit{Mbit/s},1 \textit{Mbit/s}, 256 \textit{Kbit/s}$ per \shard.
Similarly to previous experiments, we estimate processing capacity by measuring \bpt for various block sizes.
Since processing time is higher, we look at 30 seconds for our linear regression.

For~$5 \textit{Mbit/s}$, despite the limited bandwidth, the benefits of distributing transaction validation and storage outweigh the limits of cross-\shard communication.

For~$256 \textit{Kbit/s}$ per \shard, we observe a significant dip in performance when moving from a single \shard to two \shards.
This is the result of sending information across \shards becoming dominant.
Fig.~\ref{fig:intra_shard} shows these effects.
Our theoretical evaluation falls within the margin of error.
When intra-node bandwidth is extremely limited, it is more beneficial to operate a single \shard.

\subsection{Full System Simulation}
\label{sub:eval_simulation}

Finally, to assess the benefits of increasing block validation rate on a blockchain system, we take as an example the Bitcoin-NG~\cite{ng2016eyal} protocol. 
We simulate~1,000 nodes implemented with Ostraka instead of unified nodes as in the original paper.
Each node is connected to~8 other nodes at random, similar to Bitcoin.
We target 10 minutes between leader election blocks and 10 seconds between microblocks, as  Bitcoin-NG.
We produce an event-driven simulation, with identical node configurations, a network topology similar to Bitcoin, and fixed bandwidth per node.
Fig.~\ref{fig:bitcoin_ng} shows the effect of adding additional \shards to each node.
Network topology and properties strongly affect the protocol's performance, but the trend is clear. 
Ostraka allows the system to achieve an order of magnitude improvement in throughput. 

Table~\ref{table:comapre} summarizes the comparison of \sys with Bitcoin-NG, OmniLedger, and RapidChain (cf.~\S\ref{sec:related}).
The table compares results as reported by the two papers.
We compare the adversarial percentage of the network the system is resilient against.
In the case of \sys, we consider the Bitcoin-NG \pow setup we used in our simulation.
Using the Bitcoin-NG consensus algorithm, the system is resilient against a~49\% adversary.
OmniLedger and RapidChain require a high number of nodes and participants to reach high throughput and good security guarantees.
Ostraka, on the other hand, does not require a lower bound on the number of nodes.
Deployed in democratic environments, where voting power is distributed equally across the network, Omniledger and RapidChain outperform Ostraka by sharding the bandwidth and the computational resources.
However, in systems whose voting power is not uniformly distributed (e.g., Bitcoin, Ethereum), it is possible to apply the Ostraka design to reach high throughput. 

\begin{table}
 \centering
\begin{minipage}{.45\textwidth}
 \caption{Ostraka compared to other systems}
 \label{table:comapre}
 \begin{tabular}{|c | c | c  | c |}
 \hline
  & OmniLedger & RapidChain & Ostraka \\ [0.5ex]
  \hline
  \hline
 Security & 25\% & 33\% & 50\% \\
 \hline
 Cross-Shard Tx & By Client$^a$ & By leader$^b$ & Internal \\
 \hline
 Required Nodes & hundreds & hundreds & any \\
 \hline 
 Throughput & 10k tx/s$^c$ & 7.3k tx/s$^d$ & 10k tx/s$^e$ \\
 \hline
 DoS resilience & \ding{56} & \ding{56} & \ding{52} \\
 \hline
\end{tabular}
\end{minipage}
\footnotesize{$^a$Client must be online. $^b$Assuming an honest leader. $^c$400 nodes. $^d$1800 nodes, 12.5\% adversary, $^e$1000 nodes, with Bitcon-NG}

\end{table}
    \section{Conclusion}
    \label{sec:conclusion}

We present Ostraka, an architecture for scalable blockchain nodes.
Ostraka overcomes the security and performance limitations of previous solutions and achieves linear scaling in the number of node-shards.
Our experiments reached a rate of~$400k$~tx/sec at the capacity of our experimental testbed. 
Together with a high-performance consensus protocol, Ostraka can saturate network capacity. 

\section*{Acknowledgment}
This research was supported by the Israel Science Foundation (grant No.\ 1641/18), the Technion Hiroshi Fujiwara cyber-security research center, and the Israel cyber bureau.


\let\oldbibliography\thebibliography
\renewcommand{\thebibliography}[1]{%
  \oldbibliography{#1}%
  \setlength{\itemsep}{0pt}%
}

\bibliographystyle{IEEEtranS}
\bibliography{references}


\end{document}